\documentclass[aps,prb,twocolumn,floats,floatfix,showpacs,footinbib,superscriptaddress]{revtex4-1}
\pdfoutput=1
\usepackage{calc}
\usepackage{tikz}
\usepackage{graphicx}
\usepackage{color}
\usepackage{amsmath, amssymb}
\usepackage[percent]{overpic}
\usepackage{array}
\usepackage{hyperref}
\usepackage{multirow}

\newcommand{\abs}[1]{\left| #1 \right|} 
\newcommand{\avg}[1]{\left< #1 \right>} 
\newcommand{\f}[2]{\frac{#1}{#2}}

\begin{document}
\title{Anderson transition and multifractals in the spectrum of the
  Dirac operator of Quantum Chromodynamics at high temperature} 
\author{L{\'a}szl{\'o} \surname{Ujfalusi}}
\email[Contact: ]{ujfalusi@phy.bme.hu}
\affiliation{Elm{\'e}leti Fizika Tansz{\'e}k, Fizikai Int{\'e}zet, 
Budapesti M{\H{u}}szaki {\'e}s Gazdas{\'a}gtudom{\'a}nyi Egyetem, \\
H-1521 Budapest, Hungary}
\author{Matteo \surname{Giordano}}
\affiliation{Institute for Nuclear Research of the Hungarian Academy
  of Sciences,  \\
Bem t{\'e}r 18/c H-4026 Debrecen, Hungary}
\author{Ferenc \surname{Pittler}}
\affiliation{MTA-ELTE Lattice Gauge Theory Research Group,\\
  P{\'a}zm{\'a}ny P.\ s{\'e}t{\'a}ny 1/A H-1117 Budapest, Hungary}
\author{Tam{\'a}s G. \surname{Kov{\'a}cs}}
\email[Contact: ]{kgt@atomki.mta.hu}
\affiliation{Institute for Nuclear Research of the Hungarian Academy
  of Sciences,  \\
Bem t{\'e}r 18/c H-4026 Debrecen, Hungary}
\author{Imre \surname{Varga}}
\affiliation{Elm{\'e}leti Fizika Tansz{\'e}k, Fizikai Int{\'e}zet, 
Budapesti M{\H{u}}szaki {\'e}s Gazdas{\'a}gtudom{\'a}nyi Egyetem, \\
H-1521 Budapest, Hungary}
\date{\today}
\begin{abstract}
We investigate the Anderson transition found in the spectrum of the Dirac
operator of Quantum Chromodynamics (QCD) at high temperature, studying the
properties of the critical quark eigenfunctions. Applying multifractal
finite-size scaling we determine the critical point and the critical
exponent of the transition, finding agreement with previous results,
and with available results for the unitary Anderson model. We estimate
several multifractal exponents, finding also in this case agreement
with a recent determination for the unitary Anderson model. Our
results confirm the presence of a true Anderson
localization-delocalization transition in the spectrum of the quark
Dirac operator at high-temperature, and further
support that it belongs to the 3D unitary Anderson model class. 
\end{abstract}
\maketitle


\section{Introduction}
\label{sec:qcd_intro}
The Anderson metal-insulator transition is a genuine quantum
phase transition, which has been widely investigated in condensed
matter physics since the seminal paper of Anderson~\cite{Anderson58}. 
In the past years Anderson transitions were found in a wide range of
physical systems, such as ultrasound in disordered elastic
networks~\cite{Hu08,Faez09}, light in disordered photonic lattices in
the transverse direction~\cite{Segev13}, or in an ultracold atomic
system in a disordered laser trap~\cite{AspectBouyer12}.

A characteristic feature of Anderson transitions is the rich
multifractal structure of critical eigenstates, which has been the
subject of intense research activity in recent years (see
Ref.~\onlinecite{EversMirlin08} for a review). Direct signs of
multifractals at the metal-insulator transition point have been
observed experimentally in dilute magnetic
semiconductors~\cite{Richardella10}. Multifractality can moreover
influence the behavior of various systems near criticality in 
different ways. For example, the large overlap of multifractal
wave-functions can increase the superconducting critical
temperature~\cite{Feigelman10Burmistrov12}. The multifractality of the   
local density of states may induce a new phase because of the presence
of local Kondo effects induced by local pseudogaps at the Fermi
energy~\cite{Kettemann12}. 


The simplest model displaying a metal-insulator transition
is the Anderson model, which describes non-interacting
fermions in a disordered crystal. Disorder is usually introduced
through a random on-site potential, while hopping elements are fixed
(up to a random phase or $SU(2)$ rotation). In this case the system
belongs to one of the Wigner-Dyson (WD) symmetry classes depending on
the global symmetries of the system. On the other hand, systems with
vanishing on-site terms and random hopping terms, if the lattice is
bipartite, possess an additional chiral symmetry and belong to one of
the chiral WD classes~\cite{EversMirlin08}.


Quite surprisingly, an Anderson transition has been shown to take
place also in the spectrum of the Dirac operator in Quantum
Chromodynamics (QCD) at high
temperature~\cite{GarciaGarcia:2006gr,Pittler,KP,Kovacs12,Giordano14}
(see Ref.~\onlinecite{Giordano:2014qna} for a review). QCD  
is the quantum field theory governing strong interactions at the
microscopic level, and operates on length and energy scales vastly
different from the ones usually encountered in condensed matter
physics. QCD is a non Abelian gauge theory, describing the
interactions of quarks, which are fermions, and gluons, which are the
vector bosons of the $SU(3)$ gauge symmetry. Although these are the
fundamental degrees of freedom, they do not appear in the spectrum of
the theory at low temperatures, which contains only hadrons, i.e.,
bound states of quarks and gluons, due to the phenomenon of
confinement. However, at a (pseudo)critical temperature, $T_c$,
strongly interacting matter undergoes a crossover to the so-called
quark-gluon-plasma phase, and at high temperatures quarks and gluons
are deconfined. This transition is accompanied by
the restoration of the approximate chiral symmetry that is
spontaneously broken at low temperatures~\cite{Aoki:2006we}.

\begin{figure*}[th!]
  \begin {center}
  \begin{tabular}{c c c}
  \begin{overpic}[width=.30\textwidth]{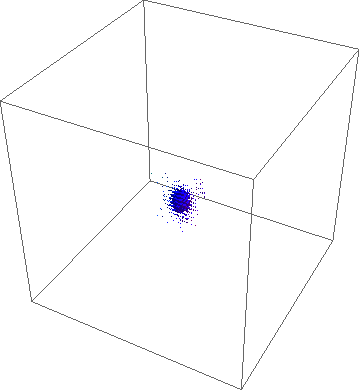}
   \put(5,95){\large (a)}\end{overpic}&
  \begin{overpic}[width=.30\textwidth]{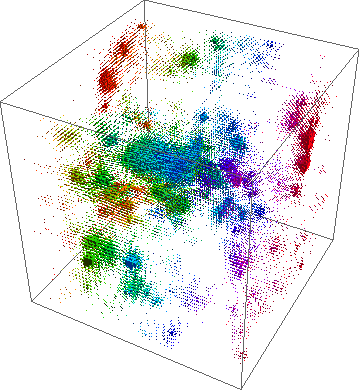}
    \put(5,95){\large (b)}\end{overpic} &
  \begin{overpic}[width=.30\textwidth]{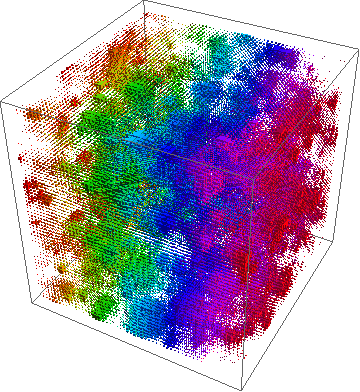}
    \put(5,95){\large (c)}\end{overpic} \\
  \multicolumn{3}{c}{\begin{tikzpicture}\draw[-] (0,-1.1) -- node[below, pos=0.5416666667] {energy} (12,-1.1); \filldraw[left color=white, right color=black]  (0,-1.1)-- (12,-1.1)--(12,-1.3)--(13,-1)--(12,-0.7)--(12,-0.9)--(0,-0.9)--cycle; \end{tikzpicture}}\\
  \end{tabular}
  \caption{
    Eigenvectors of the Dirac operator in lattice QCD at $T\approx 2.6
    T_c$ (a) in the insulating regime, (b) at
    criticality, and (c) in the metallic regime. 
    Dot sizes are proportional to
    $\sqrt{|\psi|^2}$, with proportionality factors tuned 
    independently for each subfigure to improve visualization. 
    The spatial system size is $L=56$ (in lattice units) for all
    subfigures. Coloring is 
    determined by the value of the $x$ coordinate.
  }  
  \label{fig:qcd_eigvecs}
  \end{center}
\end{figure*}
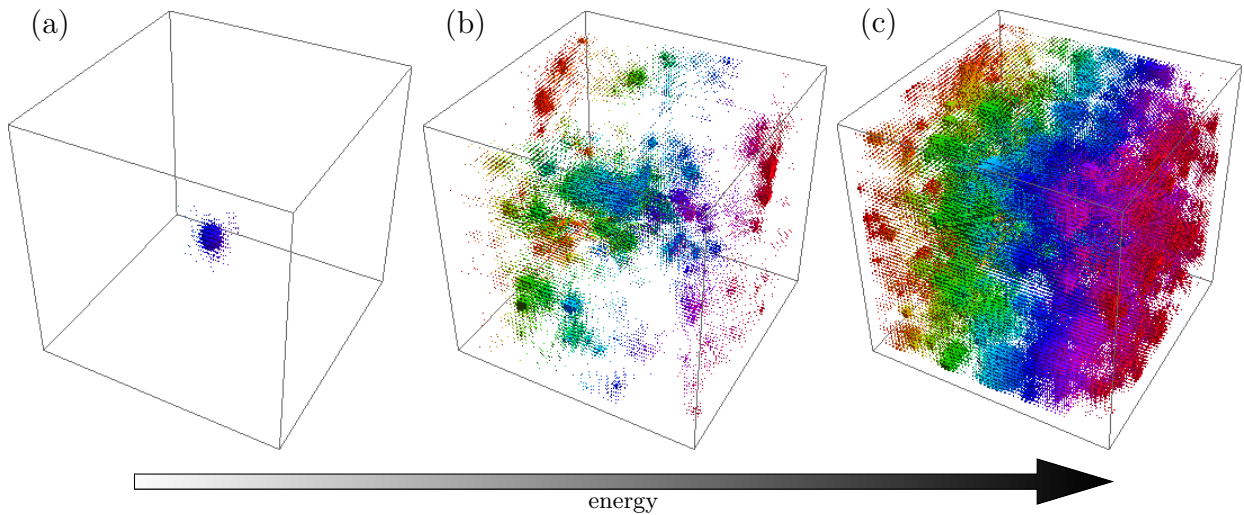

Contributions of quarks to observables, as well as all quark
correlation functions, are entirely encoded in the eigenvalues and the
eigenvectors of the Dirac operator in the background of a non Abelian
gauge field. Physical quantities are then obtained after averaging
over the gauge field configurations with the appropriate path-integral
measure. In this respect, the eigenmodes of the Dirac operator can be
formally treated 
as the eigenstates of a random
``Hamiltonian'', with disorder provided by the gauge field
fluctuations. Among the eigenmodes, a prominent role is played by the
low-lying ones: for example, they give the most important
contributions to the quark correlation functions, and determine the
fate of chiral symmetry through the Banks-Casher
relation~\cite{Banks:1979yr}. The low end of the spectrum looks
completely different in the hadronic and in the quark-gluon-plasma
phase. At low temperatures, the density of states is finite near the
origin, and both low-lying and bulk eigenmodes are extended throughout
the system. In contrast, at high temperatures, above $T_c$, the
density of states vanishes at the origin, and the low-lying eigenmodes
are localized on the scale of the inverse temperature, while higher up
in the spectrum, beyond a temperature-dependent critical ``energy'',
$E_c(T)$, the eigenmodes 
are again extended~\cite{KP,Kovacs12}. The temperature dependence of
the ``mobility edge'' was investigated in Ref.~\onlinecite{Kovacs12},
in which it was found that $E_c(T)$ extrapolates to zero at a
temperature compatible with $T_c$. Typical Dirac eigenmodes in the
localized, critical and delocalized regimes are shown in
Fig.~\ref{fig:qcd_eigvecs}. The transition in the spectrum from
localized to delocalized eigenmodes has been shown to be a
second-order phase transition, with critical exponent compatible with 
the one found in the three-dimensional unitary Anderson
model~\cite{Giordano14}.

It is rather surprising at first that the Anderson transition in the
high-temperature QCD Dirac spectrum seems to belong to the same
universality class as that of the three-dimensional unitary Anderson
model. From the point of view of statistical systems,
QCD at a finite temperature $T$ is in fact a four-dimensional Euclidean
model, with the ``temporal'' dimension compactified on a circle of
length $1/T$. However, it has been argued 
that high-temperature QCD is an 
effectively three-dimensional disordered system with on-site
disorder, the strength of which is set by the
temperature~\cite{Giordano:2015vla,Bruckmann:2011cc}.  
While this makes it more plausible that the two
models actually belong to the same universality class, it does not
make it less important to look for further evidence. In this respect,
finding the same multifractal structure in the critical eigenstates
would give strong support to the claim of
Ref.~\onlinecite{Giordano14}, and so to the broader universality of
the critical properties of Anderson transitions. 
The study of this multifractal structure is precisely the
aim of this work.  

The plan of the paper is the following. In section \ref{sec:MFSS} we
give a brief discussion of multifractality, and of the method of
multifractal finite-size scaling (MFSS). In section
\ref{sec:qcd_numerics} we 
describe in some detail the Dirac operator and the numerical
simulations of QCD employed in this paper. In section
\ref{sec:qcd_corrE} we study the correlations between eigenvectors of
the Dirac operator around the critical energy, both for comparison to
the 3D unitary Anderson model, and for their appropriate treatment in
the statistical analysis. In section \ref{sec:qcd_MFSS} we discuss the
results of MFSS for the eigenvectors of the Dirac operator. Finally,
in section \ref{sec:concl} we state our conclusions.

\section{Finite-size scaling laws for generalized multifractal
  exponents} 
\label{sec:MFSS}

In this section we briefly review wave-function multifractality and 
the technique of multifractal finite-size scaling. The wave-function
$\psi(\vec x)$ of a particle in $\mathbb{R}^d$ is naturally associated
to a local probability distribution, namely $|\psi(\vec x)|^2$,
giving the probability to find the particle in an infinitesimal
neighborhood of $\vec x$. For smooth wave functions, the probability
to find the particle in a small but finite neighborhood of $\vec x$ 
of size $r$ scales as $\sim r^d$. For fractal wave functions, this
probability scales as $\sim r^\alpha$, where $\alpha<d$ is called the
fractal dimension. For strongly fluctuating wave-functions,
however, this probability scales in general as $\sim 
r^{\alpha(\vec x)}$, with an $\vec x$-dependent power $\alpha(\vec x)$
called the local dimension. In turn, points with the same local
dimension, $\alpha(\vec x)=\alpha$, constitute a subset of
$\mathbb{R}^d$ characterized by its own fractal dimension, which
generally depends on $\alpha$. The wave-function therefore defines not 
one, but many different fractals, and is therefore said to be
multifractal. Multifractal wave-functions are strongly fluctuating on
every length-scale, and their characterization requires an infinite
number of fractal dimensions, called multifractal exponents
(MFEs). An example of a multifractal wave-function is shown in
Fig.~\ref{fig:qcd_eigvecs}(b).

Multifractality is a known feature of critical eigenfunctions at
the Anderson metal-insulator transition~\cite{EversMirlin08}, that
can be studied by means of multifractal finite-size scaling
(MFSS)~\cite{Rodriguez_prl}. In recent high-precision
calculations~\cite{Rodriguez11,Ujfalusi15,Ujfalusi14}, MFSS has been
successfully employed to determine the MFEs of critical
eigenfunctions, as well as to obtain a more precise estimate of the
critical disorder and of the critical exponents, for Anderson models
in different symmetry classes. In this work we want to perform a
similar MFSS analysis to study the Anderson
localization-delocalization transition in the spectrum of the Dirac
operator in QCD. 

In the remainder of this section we describe MFSS in
some detail. Our methods and notations are essentially the same as in 
Ref.~\onlinecite{Ujfalusi15}, to which we refer the reader for a more
detailed discussion. There is however one important difference,
concerning the way in which the transition is approached. In
Ref.~\onlinecite{Ujfalusi15} the transition was studied by looking at
wave functions at the band center and varying the amount of
disorder, $W$. In QCD the amount of disorder is effectively set by the
temperature, and it is more convenient to keep it fixed
and study the transition as a function of energy, $E$, by looking at
wave-functions near the mobility edge, $E_c$. Therefore,  $W$ has been
replaced by $E$ in the expressions of Ref.~\onlinecite{Ujfalusi15}.

Let us consider a $d$-dimensional cubic lattice of linear size
$L$, and a critical eigenfunction of a random Hamiltonian, $\psi(\vec
x)$, defined on the lattice sites $\vec x$ and normalized to 1. We can
divide the lattice into smaller boxes of linear size $\ell$, and
compute the probability corresponding to the $k$-th box as
\begin{equation}  \label{eq:multifractal_mu}
\mu_k=\sum_{\vec x \in {\rm box}_k} |\psi(\vec x)|^2,
\end{equation}
where the sum runs over the lattice sites contained in the $k$-th
box. The generalized inverse participation ratios (GIPRs) are the
moments of the box probability. 
The GIPRs and their derivatives read
\begin{equation}\label{eq:multifractals_SqRq} 
R_q=\sum_{k=1}^{\lambda^{-d}} \mu_k^q\qquad 
S_q=\frac{dR_q}{dq}=\sum_{k=1}^{\lambda^{-d}} \mu_k^q \ln \mu_k,
\end{equation}
where $\lambda=\frac{\ell}{L}$, and the sum runs over all the
$\lambda^{-d}$ boxes of size $\ell$. For small $\lambda$, 
the averages of $R_q$ and $S_q$ over disorder realizations follow a
power-law behavior as a function of $\lambda$, which leads one to define
the following exponents: 
\begin{equation} 
D_q=\lim_{\lambda\to 0}\frac{1}{q-1}\frac{\ln \avg{R_q}}{\ln
  \lambda}\qquad \alpha_q=\lim_{\lambda\to
  0}\frac{\avg{S_q}}{\avg{R_q}\ln\lambda}.
\label{eq:multifractals_D_alpha}\end{equation}
$D_q$ and $\alpha_q$ are generalized fractal dimensions, usually
referred to as multifractal exponents (MFEs). 
One can similarly define MFEs for localized and delocalized states by 
substituting critical eigenfunctions with localized or delocalized
eigenfunctions in Eq.~\eqref{eq:multifractal_mu}. In the
delocalized/metallic part of the spectrum, states extend over the
whole lattice, so their effective size grows proportionally to
the volume, thus leading to $D_q^{met}\equiv d$. On the other hand, in
the localized/insulating regime, states are exponentially localized, so
that their effective size does not change with the system size,
resulting in $D_q^{ins}\equiv 0$ for $q> 0$, and $D_q^{ins}\equiv
\infty$ for $q<0$. At criticality, $E=E_c$, the eigenstates are
instead expected to be multifractal, with nontrivial, $q$-dependent
$D_q$ and $\alpha_q$.  

This jump of the MFEs at the critical point happens only in an
infinite system. The main idea of MFSS is to use the MFEs as order
parameters for finite size-scaling. In order to do that we have to
define the finite size version of the MFEs at a given energy, 
\begin{eqnarray}
\label{eq:alphaD_ens}
\tilde{\alpha}_q^{ens}(E,L,\ell) &=& \frac{\avg{S_q}}{\avg{R_q}\ln\lambda}\,,\\
\tilde{D}_q^{ens}(E,L,\ell) &=& \frac{1}{q-1}\frac{\ln\avg{R_q}}{\ln\lambda}\,,
\end{eqnarray}
where it is understood that wave-functions of energy around $E$ are
used on the right-hand side, and where the
superscript {\it ens} is to remind the reader that one has to perform 
{\it ensemble} averaging over the different disorder realizations. 
$\tilde{\alpha}_q$ and $\tilde{D}_q$ are called generalized
multifractal exponents (GMFEs). Every GMFE approaches the value of the
corresponding MFE at the critical point, $E=E_c$, only in the limit
$\lambda\to 0$.
One can also define {\it typical} MFEs, 
\begin{eqnarray} 
\tilde{\alpha}_q^{typ}(E,L,\ell) &=&
\avg{\frac{S_q}{R_q}}\frac{1}{\ln\lambda}\,,\\ 
\tilde{D}_q^{typ}(E,L,\ell) &=& \frac{1}{q-1}\frac{\avg{\ln
    R_q}}{\ln\lambda}\,, 
\label{eq:alphaD_typ}
\end{eqnarray} 
which can be used as well in a finite-size scaling analysis.
However, 
as we said above, MFEs are defined
through {\it ensemble} averaging in principle [see
Eq.~(\ref{eq:multifractals_D_alpha})], and when computing MFEs in
Sec.~\ref{sec:qcd_MFSS} we use {\it ensemble} averaged quantities
only.  

In the renormalization group language, the Anderson transition is
characterized by a single relevant operator~\cite{AALR}, and so in the
vicinity of the critical point one can derive  
scaling laws for the GMFEs,
which can be summarized in a single equation, using a common
letter, $G$, for the GMFEs: 
\begin{equation} \tilde{G}_q(E,L,\ell) =
  G_q+\frac{1}{\ln\lambda}\mathcal{G}_q\left(\frac{L}{\xi},
\frac{\ell}{\xi}\right) \,.
\label{eq:fss_scalinglaw_Ll}
\end{equation}
At the critical point, the localization length diverges as
$\xi\sim[\varrho(E-E_c)]^{-\nu}$, where $\varrho(E-E_c)\approx E-E_c$
for $(E-E_c)/E_c\ll 1$. The system sizes employed in this paper,
however, are not big enough to justify the use of one-parameter
scaling, and so we included the contribution of an irrelevant
operator, $\eta=\eta(E-E_c)$, which leads us to write
\begin{eqnarray} 
\tilde{G}_q(E,L,\ell) =
G_q &+& \frac{1}{\ln\lambda} \left[\mathcal{G}^{r}_q\left(\varrho
    L^{\frac{1}{\nu}}, \varrho
    \ell^{\frac{1}{\nu}}\right)+\right. \nonumber\\ 
 &+&\left. \eta \ell^{-y}\mathcal{G}^{ir}_q\left(\varrho
     L^{\frac{1}{\nu}},\varrho \ell^{\frac{1}{\nu}}\right)\right].
\label{eq:fss_anderson_scalinglaw_Ll}
\end{eqnarray}
A second irrelevant term, proportional to $L^{-y'}$, is expected to be
less important and will be neglected in the
analysis~\cite{Rodriguez11,Ujfalusi15}. 

Fits to the numerical data are performed by expanding 
$\mathcal{G}^{r}$ and $\mathcal{G}^{ir}$ in the variables $\varrho
L^{\frac{1}{\nu}}$ and $\varrho \ell^{\frac{1}{\nu}}$ up to order
$n_{r}$ and $n_{ir}$, respectively. The number of parameters therefore
grows as $\sim n_{r}^2+n_{ir}^2$. Moreover, $\varrho$ and $\eta$
must also be expanded in powers of $E-E_c$ up to order $n_\varrho$
and $n_\eta$, which further increases the number of fitting parameters.
The fit provides all the physically interesting quantities, namely the 
critical point, $E_c$, the critical exponent, $\nu$, the irrelevant
exponent $y$, and the MFE, $G_q$. 

A simpler fit can be performed by setting $\lambda=\ell/L$ to a fixed
value. In this case, dropping $\lambda$ from the notation, we can
write 
\begin{equation} 
     \tilde{G}_q(E,L) = {\cal G}_q\left(\frac{L}{\xi}\right) =
     {\cal G}^{r}_q\left(\varrho L^{\frac{1}{\nu}}\right) + \eta
     L^{-y}{\cal G}^{ir}_q\left(\varrho L^{\frac{1}{\nu}}\right), 
\label{eq:fss_anderson_scalinglaw_lambda}
\end{equation}
having absorbed $G_q$ and the factor $\lambda^{-y}/\ln \lambda$ into new
functions ${\cal G}_q^{r/ir}$. 
The main advantage is that since ${\cal G}_q^{r/ir}$ are now
single-variable functions, the number of expansion parameters grows
only as $\sim n_{r}+n_{ir}$. On the other hand, with this method one
can determine only $E_c$, $\nu$, and $y$, while the value of the MFE,
$G_q$, cannot be obtained.

\section{Properties of the Dirac operator and details of the
  simulations} 
\label{sec:qcd_numerics}

In this section we give the relevant details about the Dirac
operator and QCD, and about how the QCD Dirac spectrum can be studied
by means of numerical simulations. The continuum Euclidean Dirac operator is
\begin{equation}
D(A)=\sum_{\mu=1}^4 \gamma_\mu(\partial_\mu+igA_\mu),
\end{equation}
where $\gamma_\mu$ are the Euclidean Dirac matrices,
$g$ is the coupling constant, and $A_\mu$ is the non Abelian gauge
field. More precisely, $A_\mu=\sum_a A_\mu^a t^a$ is a Hermitian
$3\times 3$ matrix, where $A_\mu^a=A_\mu^a(x)=A_\mu^a(\vec {x},t)$ is
real and the sum runs over the generators 
$t^a$ of $SU(3)$. The Dirac operator is thus anti-Hermitian, so
admitting a straightforward interpretation as ($i$ times) the
Hamiltonian of a quantum system. The Dirac operator is a chiral
operator with the following structure in spinor space, 
\begin{equation}
D=
i \left(\begin{matrix}
0 &W\\
W^\dagger & 0\\
\end{matrix}\right),
\end{equation}
with $W$ a complex matrix with no further
symmetry~\cite{Verbaarschot00}. As a random matrix model, the Dirac
operator in a random gauge field belongs therefore to the chiral
unitary class. Chiral symmetry is expressed by the anticommutation
relation $\{\gamma_5,D\}=0$, which implies that the nonzero
eigenvalues come in pairs $\pm iE_n$. It is thus sufficient to
consider the positive part of the spectrum only.

The partition function of QCD at temperature $T$ can be expressed
as a functional integral,
\begin{equation}
  \label{eq:qcd_part}
  Z_{\rm QCD} = \int [dA]\,e^{-S_{\rm g}[A]} \prod_f\det [D(A) + m_f] \,,
\end{equation}
with the constraint $A_\mu(\vec {x},1/T)=A_\mu(\vec {x},0)$. The
product is over the six different types of quarks
(``f\mbox{}lavors''), with $m_f$ the mass of quark $f$. 
Here $S_{\rm g}[A]$ is a positive functional of the gauge field,
which together with the determinants provides the probability
distribution of the disorder, i.e., of the gauge field
configurations. Numerical simulations of QCD require the
discretization of Eq.~\eqref{eq:qcd_part} on a finite lattice.
For a review of lattice QCD see, e.g., Ref.~\onlinecite{MontvayMunster}. 
While the discretization of the gauge fields poses no particular
problem, and can be performed 
preserving exact gauge invariance~\cite{Wilson:1974sk}, fermion fields
are known to be more problematic, and the discretization of the Dirac
operator spoils some of the properties of its continuum counterpart.  
Nevertheless, the discretization that we employed, namely staggered
fermions~\cite{Susskind:1976jm}, preserves the anti-Hermiticity and
the symmetry of the spectrum with respect to the origin, and moreover
preserves the chiral unitary symmetry class~\cite{Verbaarschot00}.

It must be noted at this point that in the case of the Anderson model,
chiral and non-chiral symmetry classes differ only in their
properties near the band 
center~\cite{Evangelou03}, i.e., $E=0$, while the properties of the
bulk of the spectrum are similar. For example, the authors of
Ref.~\onlinecite{Evangelou03} found Wigner-Dyson statistics 
in the bulk spectrum of a three-dimensional chiral orthogonal
disordered model. Moreover, even the critical exponent of the
orthogonal and of the chiral orthogonal class turn out to be the same,
up to very high numerical precision~\cite{Biswas00}. 
We expect the same to be true for the multifractal
  exponents.

Let us now describe the numerical setting in some detail. QCD is
discretized on a periodic hypercubic lattice $x_\mu\in\mathbb{Z}$, of
spatial extent $L$ in each direction and temporal extent $L_t$. The
gauge fields $A_\mu$ are replaced by corresponding gauge links, i.e.,
parallel transporters along each link of the lattice, which are
elements of the gauge group, $SU(3)$. The functional $S_{\rm g}$ is
discretized and expressed in terms of the gauge links, and the
integration over gauge fields is replaced by the integration with the
Haar measure over gauge links, i.e., over the gauge-group valued
variables on the links. Finally, the continuum Dirac operator is
replaced by the staggered Dirac operator, which reads
\begin{equation}
  \label{eq:stag_op}
  D^{\rm stag}_{xy} = \f{1}{2}\sum_{\mu=1}^4 \eta_\mu(x)
\left[  \delta_{x+\hat\mu,y}U_\mu(x)
-  \delta_{x-\hat\mu,y}U_\mu^\dag(x-\hat\mu)\right]\,,
\end{equation}
with $\eta_\mu(x)=(-1)^{\sum_{\nu<\mu} x_\nu}$, and
$U_\mu(x)\in SU(3)$ the gauge link connecting the lattice
site $x$ to the neighboring site along direction $\hat\mu$. The
staggered Dirac operator carries only spacetime and color indices,
i.e., it has no spinorial structure. The eigenvalue equation $ D^{\rm
  stag}\chi = iE\chi$ must be supplemented with the antiperiodic
boundary condition $\chi(\vec x,L_t) = -\chi(\vec x,0)$ for the quark
eigenfunction. 

As we have already remarked, the Dirac operator can be viewed as a
random Hamiltonian, with disorder provided by the fluctuations of the
gauge fields, and distributed according to the Boltzmann weight
appearing in the partition function. In its discretized version, the
Dirac operator is a large sparse matrix, with nonzero random elements
only in the off-diagonal, nearest-neighbor hopping terms, which depend
on the parallel transporter on the corresponding link of the
lattice. This resembles an Anderson model with off-diagonal disorder,
although here the fluctuations of the gauge links are correlated,
rather than independent. However, since the theory has a mass gap,
correlations decrease exponentially with the distance. 
Moreover, the strong correlation between the different time-slices
makes the model effectively three-dimensional, with the fluctuations
of the temporal links acting effectively as a three-dimensional
diagonal disorder~\cite{Bruckmann:2011cc,Giordano:2015vla}.
The size of the gauge field fluctuations are determined by the
temperature, which therefore is expected to play the same role as the
amount of disorder in the Anderson model. This is confirmed by the
fact that the temperature governs the position of the mobility edge.

In the present work we have studied the spectrum of the Dirac operator
by generating gauge link configurations, i.e., realizations of
disorder, by means of Monte-Carlo methods. Numerical calculations were
done on a GPU cluster. In our simulations we have included only the
three lightest flavors (up, down, and strange), with  equal masses for
the up and down quark. For many purposes, this is a good approximation
of the real world. The lattice spacing in physical units was set to
$a=0.125~{\rm fm}$ and the temporal size was fixed to $L_t=4$,
resulting in the temperature $T\approx 2.6\, T_c$, well above
the crossover temperature (see
Refs.~\onlinecite{Pittler,KP,Kovacs12,Giordano14} for more details).   
Technical details about the numerical implementation and the
scale-setting procedure can be found in
Refs.~\onlinecite{Aoki:2005vt,Borsanyi:2010cj}. We have  
computed the eigenpairs of the Dirac operator from the smallest
eigenvalue up to the upper end of 
the critical region, on lattices of spatial sizes in the range
$L=24-56$ (in lattice units). A detailed list is reported in
Tab.~\ref{tab:qcd_fss_systemsize} along with the 
corresponding number of samples. 

The three-dimensional box probability, Eq.~\eqref{eq:multifractal_mu},
required for the multifractal analysis, was constructed as
follows. To have a gauge-invariant description we summed over the color
components, labelled by $c$.   
Moreover, due to the strong correlation between the lattice
time-slices, the eigenvectors of the Dirac operator look qualitatively
the same on each of them, so we can also sum 
over the time-slices, $t$. The squared amplitude $|\psi(\vec
  x)|^2$ is then defined as $|\psi(\vec x)|^2 \equiv \sum_{t,c}
|\chi_c(\vec x,t)|^2$, and provides the basic three-dimensional
spatial probability distribution, from which the box probability
distribution is then obtained in the usual way.

\begin{table}
	\begin {center}
	\begin{tabular}{|c| c|}
	\hline
	system size $(L)$ & number of samples\\ \hline
	24 & 41517 \\ \hline
	28 & 20548 \\ \hline
	32 & 19250 \\ \hline
	36 & 14869 \\ \hline
	40 & 8812 \\ \hline
	44 & 5242 \\ \hline
	48 & 7008 \\ \hline
	56 & 3107 \\ \hline
	\end{tabular}
	\caption{System sizes and corresponding number of gauge
          configurations used in this work.}
	\label{tab:qcd_fss_systemsize}
	\end{center}
\end{table}

\section{Correlations between eigenvectors}
\label{sec:qcd_corrE}

In this section we investigate the correlations between
different eigenvectors of the Dirac operator in a given gauge
configuration. Our motivation is twofold. On the one hand, we want to
compare the eigenvector correlations in QCD with the ones
found in the unitary Anderson model. On the other hand, these
correlations have to be properly taken into account when fitting the
numerical data to determine the various critical quantities, as we do
in Sec.~\ref{sec:qcd_MFSS}.

\begin{figure}
  \begin {center}
  \begin{tabular}{c c}
  \begin{overpic}[type=pdf,ext=.pdf,read=.pdf,width=.48\columnwidth]{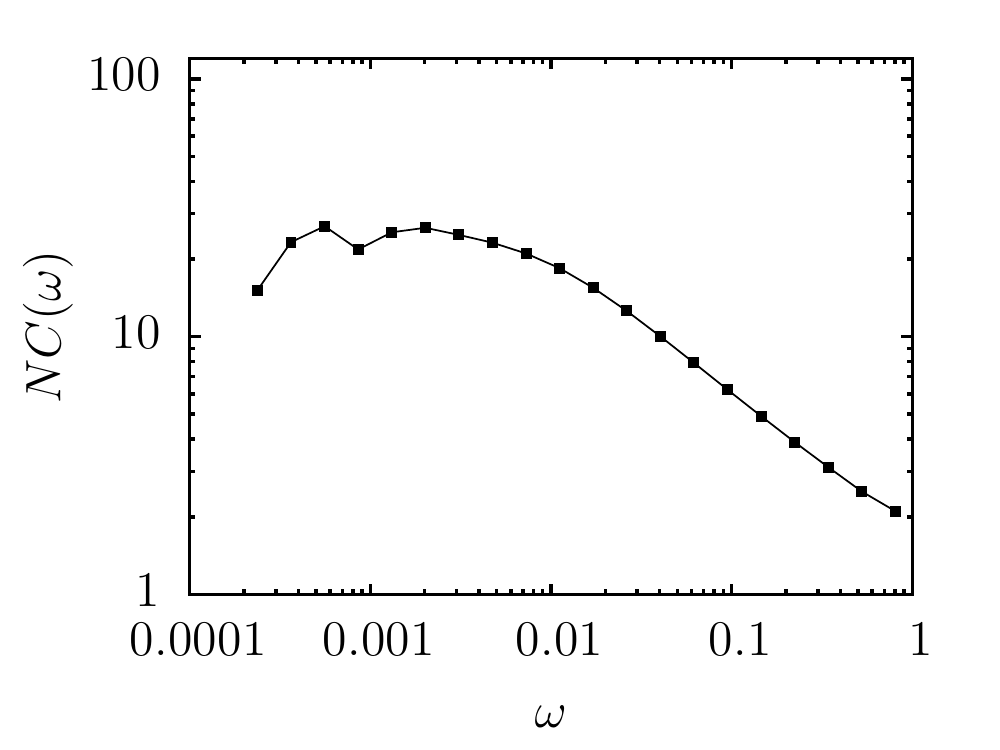} \put(-2,72){(a)}\end{overpic}
  \begin{overpic}[type=pdf,ext=.pdf,read=.pdf,width=.48\columnwidth]{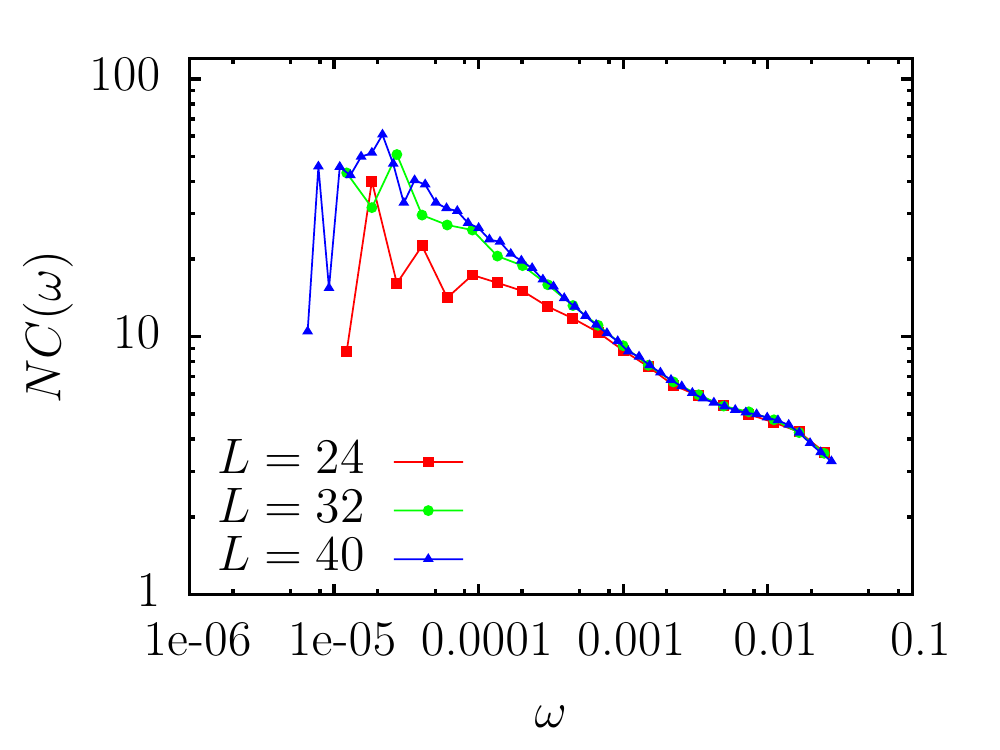} \put(-2,72){(b)}\end{overpic}
  \end{tabular}
  \caption{Correlations, Eq.~\eqref{eq:correlations}, between (a)
    critical eigenfunctions of the 
    unitary Anderson model for $W=18.37$ and system size $L=10$,
    and (b) eigenfunctions of the QCD Dirac operator at $T\approx
    2.6\,T_c$ in the critical regime, $0.32\leq 
    E\leq 0.35$, for different system sizes.} 
  \label{fig:qcd_corrE} 
  \end{center}
\end{figure}

Cuevas and Kravtsov~\cite{CuevasKravtsov07} showed that in the
Anderson model there are non-negligible correlations between
eigenvectors. Similar correlations are therefore expected also in
other disordered systems, like the one under consideration.
The relevant quantities are the density-density correlations, which
are defined in terms of the overlap
integral, which for the $i$-th and $j$-th eigenfunctions reads 
\begin{equation}\label{eq:qcd_K2}
K_2^{ij}=\int d^3x \abs{\psi_i}^2  \abs{\psi_j}^2.
\end{equation}
In the case of QCD, $\abs{\psi_i}^2$ has the meaning discussed above
at the end of Sec.~\ref{sec:qcd_numerics}. 
One then defines the joint probability
distribution of $K_2^{ij}$ and of the energy difference between
eigenstates,  
\begin{equation}
P(\omega,k)=\avg{\sum_{i,j}\delta(E_i-E_j-\omega)\delta(K_2^{ij}-k)}.
\end{equation}
To characterize the average behavior of the overlap integral as a
function of energy, its conditional expectation value is the natural
choice,  
\begin{equation}
\label{eq:correlations}
C(\omega)= 
\f{\int dk\,kP(\omega,k)}{\int dk\,
P(\omega,k)}.
\end{equation}
The quantity $C(\omega)$ is expected to be of order $1/N$ along the
whole spectrum, where $N=L^3$ is the volume of the system. Indeed, for
two delocalized states $K_2^{ij}\approx 1/N$, while for two localized
states $K_2^{ij}$ is nonzero only if they happen to be in the same
region, in which case it is of order 1, and the probability that this
happens is of order $1/N$. 

Fig.~\ref{fig:qcd_corrE}(a)
shows the eigenvector correlation $C(\omega)$ in the unitary Anderson
model at criticality. One can see a large enhancement of the
correlation at small $\omega$, and decreasing behavior with growing
energy separations, which is similar to the results of
Ref.~\onlinecite{CuevasKravtsov07} for the orthogonal Anderson
model. Examining the same correlation for critical eigenfunctions
in QCD, we also find an enhancement at small energy separations, see
Fig.~\ref{fig:qcd_corrE}(b). In the critical regime the behavior of
the two systems is very similar, and even the approximate exponent of the
power-law decay is close to $0.5$ in both cases. This is a nice
example of the similarity of the two models, and in
Sec.~\ref{sec:qcd_MFSS} we present further similarities in more detail.

\section{MFSS for the eigenvectors of the Dirac operator}
\label{sec:qcd_MFSS}
In this section we would like to characterize the Anderson phase
transition in the spectrum of the Dirac operator of QCD in the frame
of the MFSS, described in Sec.~\ref{sec:MFSS}. As discussed at the end
of Sec.~\ref{sec:qcd_numerics}, a three-dimensional spatial probability
distribution was calculated from the eigenvectors. From that, the GMFEs
$\tilde{\alpha}_q$ and $\tilde{D}_q$ were then computed according to
Eqs.~(\ref{eq:alphaD_ens})--(\ref{eq:alphaD_typ}). 
More precisely, we chose $26$ values of energy, $E_i$, in the range
$E\in[0.32,0.35]$, and for the $i$-th energy value and the $k$-th
gauge configuration we computed $R_{qi}^k$ and $S_{qi}^k$ according to
Eq.~\eqref{eq:multifractals_SqRq}. In order to decrease the numerical
noise we averaged over all the eigenvectors in an energy range of
width $\Delta E=0.0012$ around $E_i$. The GMFEs
$\tilde{\alpha}_q(E_i,L,\ell)$ and $\tilde{D}_q(E_i,L,\ell)$ are then
obtained by averaging  $R_{qi}^k$ and $S_{qi}^k$ over the index $k$,
i.e., over configurations, or in other words, over different
realizations of disorder. 

\begin{figure*}
  \begin {center}
  \begin{tabular}{c c}
  \begin{overpic}[type=pdf,ext=.pdf,read=.pdf,width=.45\linewidth]{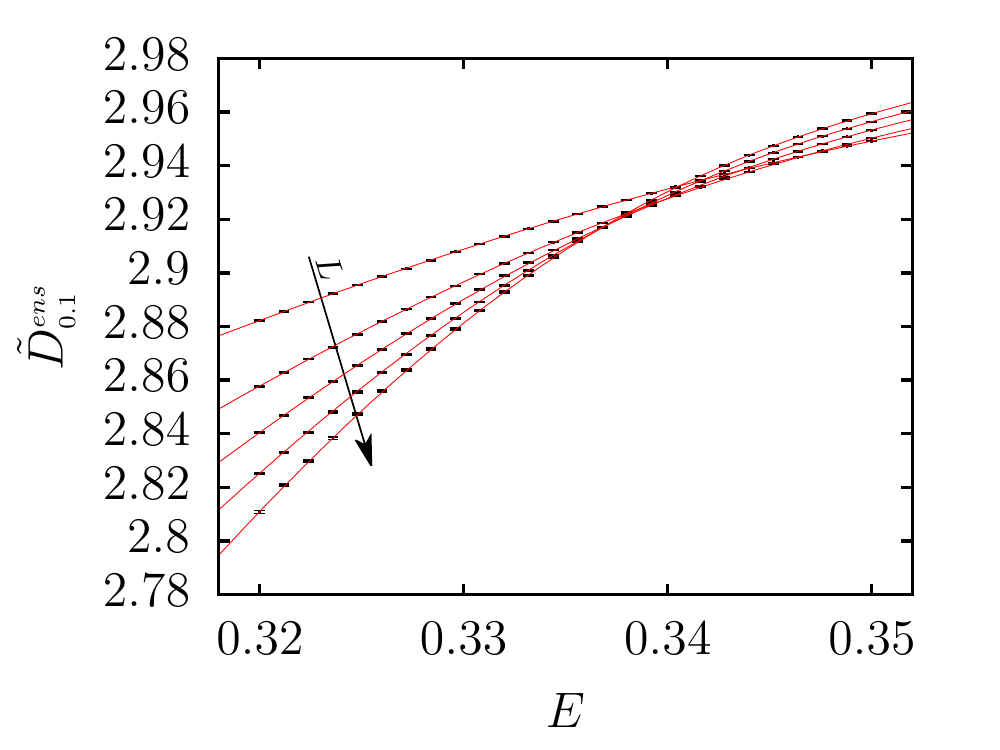}
		\put(52,18){\includegraphics[type=pdf,ext=.pdf,read=.pdf,width=.18\linewidth]{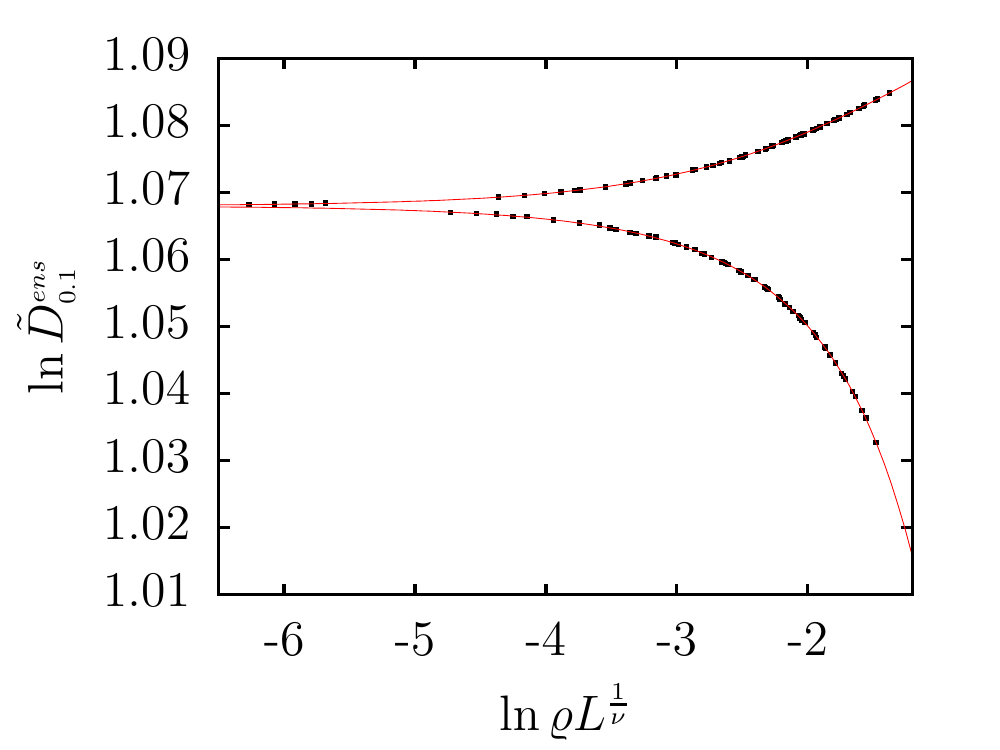}
                \put(-91,56){\large (a)}              }
	\end{overpic} &
  \begin{overpic}[type=pdf,ext=.pdf,read=.pdf,width=.45\linewidth]{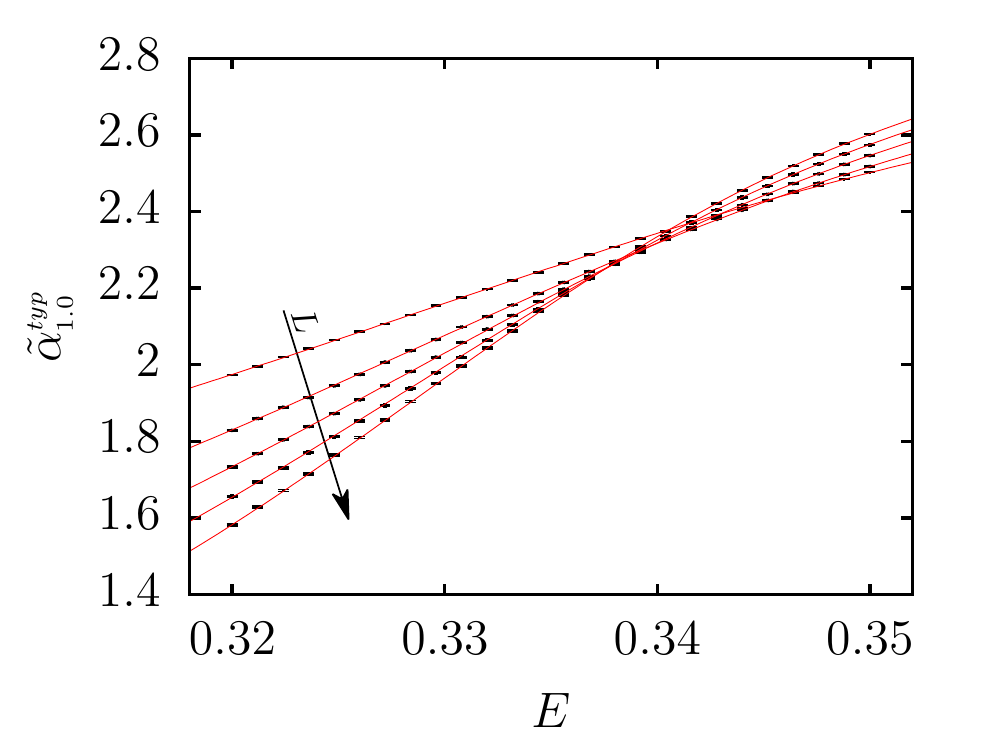} 
		\put(52,18){\includegraphics[type=pdf,ext=.pdf,read=.pdf,width=.18\linewidth]{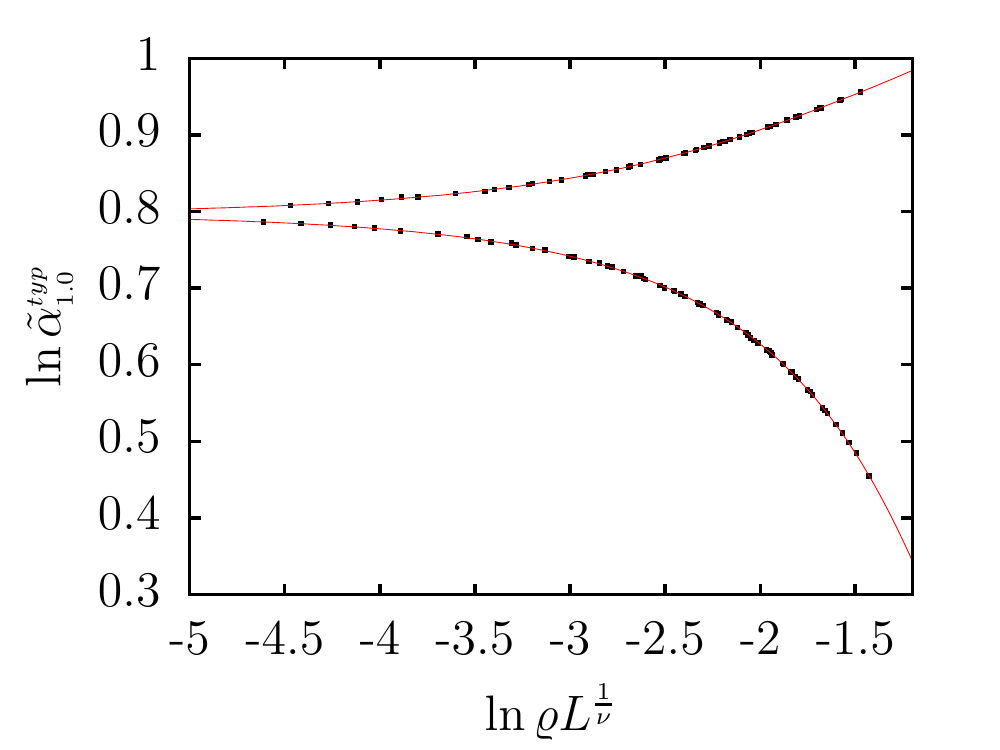}
                \put(-91,56){\large (b)}              } 
	\end{overpic}\\
  \end{tabular}
  \caption{GMFEs, (a) $\tilde{D}_{0.1}^{ens}$ and (b)
    $\tilde{\alpha}_{1.0}^{typ}$, at fixed $\lambda=0.125$.
    Dots are the raw data, and the solid red line is the best fit
    obtained by MFSS. Insets show the scaling functions on a log-log
    scale, after subtracting the irrelevant term. 
    Error bars are not shown in the insets for visual clarity.
} 
  \label{fig:qcd_lambda0125_alphaqDq_raw}
  \end{center}
\end{figure*}

An example of the resulting GMFEs at fixed $\lambda=0.125$ is depicted
in Fig.~\ref{fig:qcd_lambda0125_alphaqDq_raw}. As the system size
grows, the curves shift to opposite directions on the two sides of 
the transition. At low energy they shift down, indicating a localized
phase, while at high energy they shift up, suggesting a metallic
phase, as expected. In between, the curves should cross at a common
point, corresponding to the critical energy, but due to finite size
effects originating from the irrelevant terms this is
true only approximately. 

Data were then fitted with the scaling laws
Eqs.~(\ref{eq:fss_anderson_scalinglaw_Ll}) and
(\ref{eq:fss_anderson_scalinglaw_lambda}),  
minimizing the quantity $\chi^2/(N_{df}-1)$,
using the MINUIT library~\cite{James75}. Here $N_{df}$ is the number
of degrees of freedom and $\chi^2$ is the 
distance between the numerical data, $y_i$,
and the fitting function, $f_i$, in the appropriate metric, i.e.,
\begin{equation}
  \label{eq:chisqu}
  \chi^2 = \sum_{i,j} (y_i-f_i) (C^{-1})_{ij} (y_j-f_j),
\end{equation}
where $C$ is the covariance matrix of the data points. 
In the light of
the results of Sec.~\ref{sec:qcd_corrE}, which show that there are
strong correlations between eigenvectors in a given gauge
configuration, strong
correlations are also expected among GMFEs at different energies, and so
the inclusion of correlations in the fitting procedure 
is necessary to obtain accurate results.
The error bars of the best fit parameters were estimated by
Monte-Carlo simulation, generating $N_{\rm MC}=100$ sets of synthetic
data, distributed according to a Gaussian
  distribution with means equal to the raw data points and covariance
  matrix equal to the covariance matrix of the sample. We then
determined the error bars from the distributions of the resulting fit
parameters, choosing the $95\%$ confidence level. 

In order to perform best fits, the scaling laws
Eqs.~(\ref{eq:fss_anderson_scalinglaw_Ll}) and
(\ref{eq:fss_anderson_scalinglaw_lambda}) need to be expanded in
powers of $E-E_c$, and this requires to set the expansion orders
$n_{r/ir}$ of the relevant/irrelevant scaling term
$\mathcal{G}^{r/ir}_q$, as well as the expansion orders $n_{\varrho}$
and $n_{\eta}$ of $\varrho$ and $\eta$. Since the relevant operator is
more important than the irrelevant one we always used $n_{r}\geq n_{ir}$
and $n_{\varrho}\geq n_{\eta}$. We then repeated the fit for several
choices of the expansion orders.

The quality of the best fits was judged according to
two criteria. The first criterion was how close the ratio
$\chi^2/(N_{df}-1)$ approached unity, and only fits with
$\chi^2/(N_{df}-1)\approx 1$ were considered acceptable. The second
criterion was stability against changing the expansion orders, in
order to keep under control the systematic effects due to the
truncation of the scaling function. 
We estimated the systematic error due to truncation as twice the
standard deviation of the critical parameters, in the sample comprising
the stable fits and the essentially equivalent ones obtained by
increasing or lowering the expansion orders. The factor of 2 is
required by consistency with the 95\% confidence level chosen for the
statistical error.

We first performed the MFSS at fixed $\lambda$, as described in
Sec.~\ref{sec:MFSS}, both for ensemble and typical averaging. We used
$\lambda=0.125$, as this value is compatible with several of 
the system sizes listed in Tab.~\ref{tab:qcd_fss_systemsize}. 
The fixed $\lambda$ method is more stable, since the number of
parameters to fit grows only linearly with the expansion
orders. Stability was a serious issue, because the largest system size
available, $L=56$, was only about half of the one used in
Refs.~\onlinecite{Rodriguez11,Ujfalusi15,Ujfalusi14}.  
Due to this limitation, fits were stable for adding or removing an
expansion parameter only in the range $0\leq q \leq 1$.
The reason for this is that, for large $|q|$, the $q$-th power in
Eq.~(\ref{eq:multifractals_SqRq}) strongly enhances the contribution
of the few spatial points with very large (if $q>0$) or very small (if
$q<0$) wave-function amplitude squared, which therefore dominate the
sum. This results in an effectively reduced statistics and so in a
noisy dataset, and leads to a regime $0\leq q \leq 1$, where GMFEs
behave numerically the best. Notice that, by construction, $D_0=d=3$
and $D_1=\alpha_1$, and moreover $\alpha_{0.5}=d$ due to a symmetry
relation derived in Ref.~\onlinecite{MFME}. 

\begin{figure*}
  \begin {center}
  \begin{tabular}{c c c}
  \includegraphics[width=.325\textwidth]{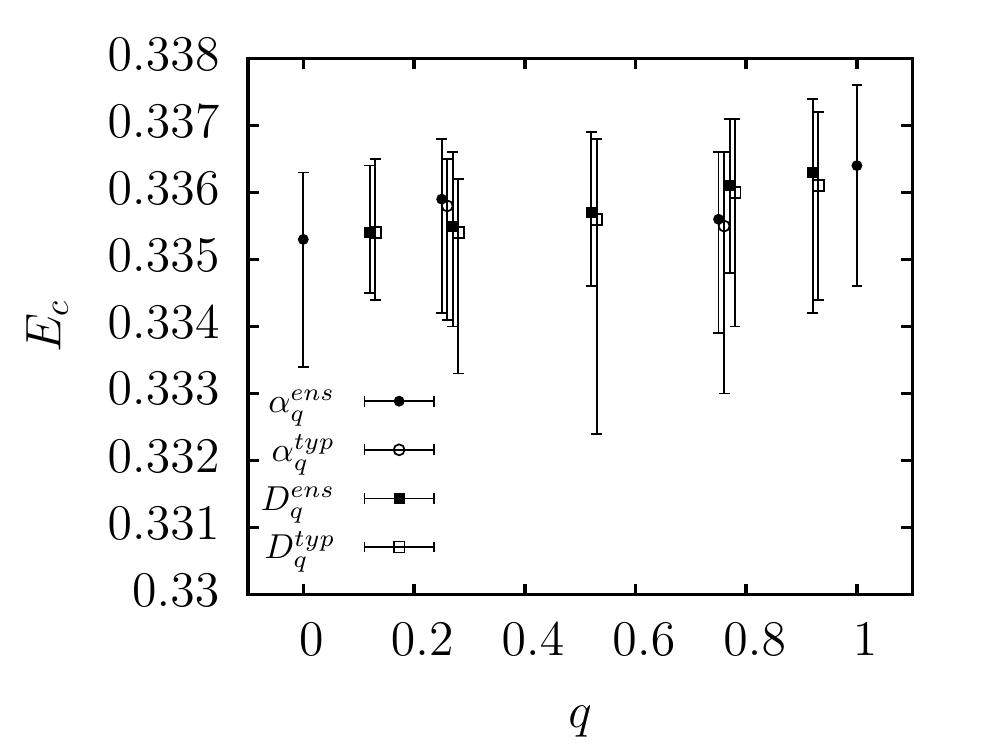} &
  \includegraphics[width=.325\textwidth]{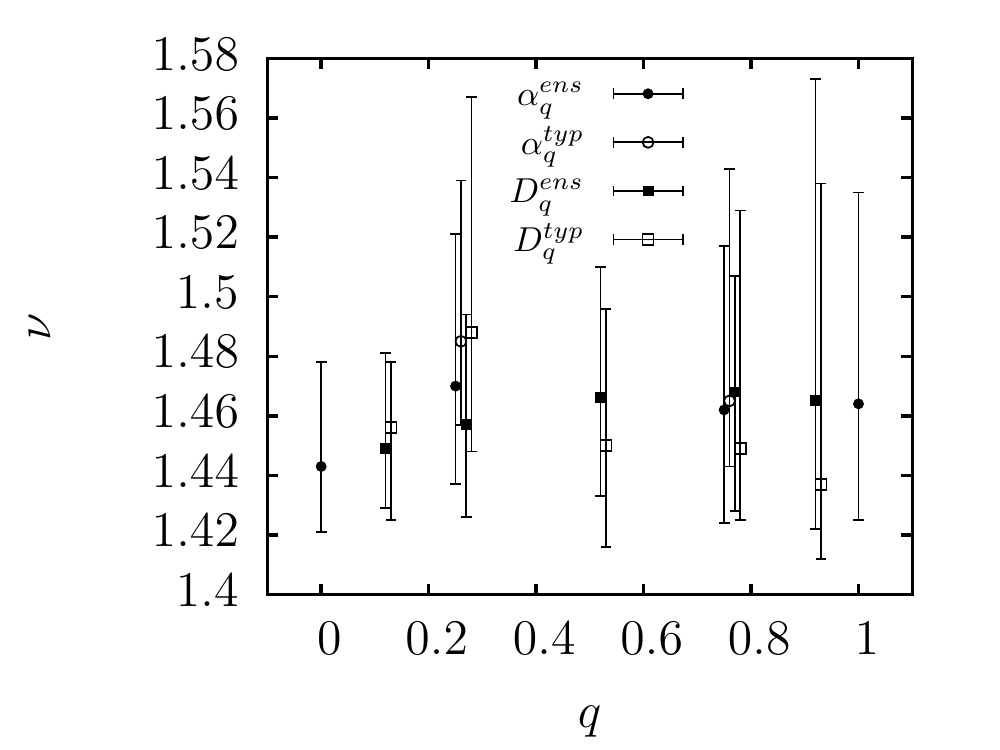} &
  \includegraphics[width=.325\textwidth]{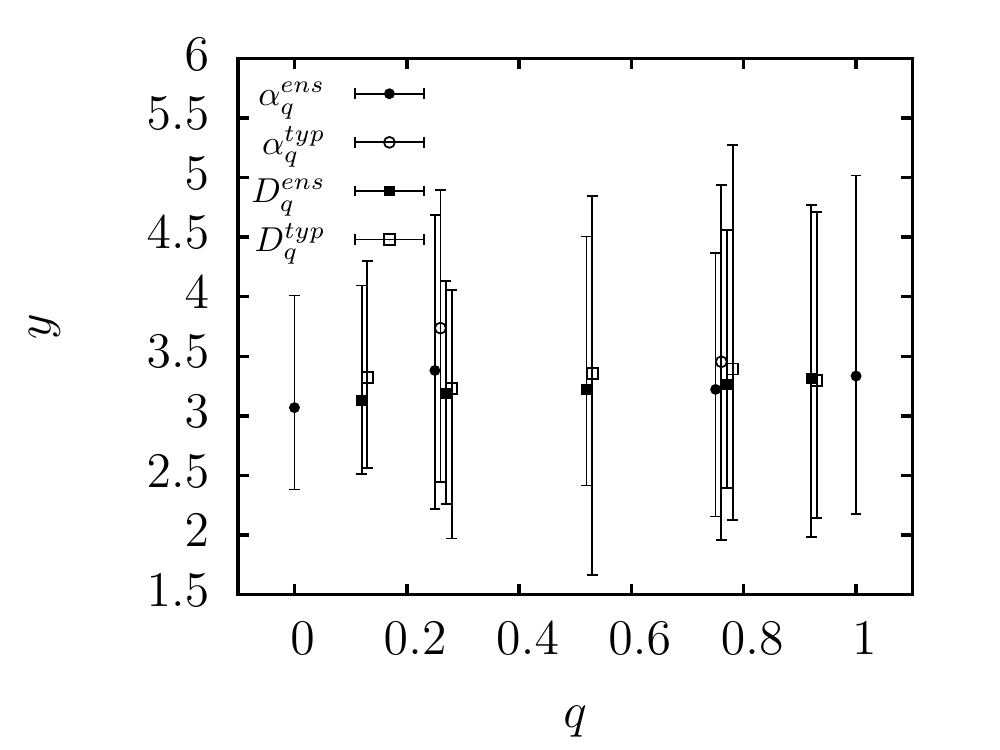} \\
  \end{tabular}
  \caption{Critical parameters obtained via MFFS at fixed
    $\lambda=0.125$ on the eigenvectors of the QCD Dirac
    operator. Error bars correspond to the $95\%$ confidence
    band. Systematic errors are not included.} 
  \label{fig:qcd_fitres_lambda0125}
  \end{center}
\end{figure*}

\begin{figure}[h!]
  \begin {center}
  \includegraphics[width=.49\columnwidth]{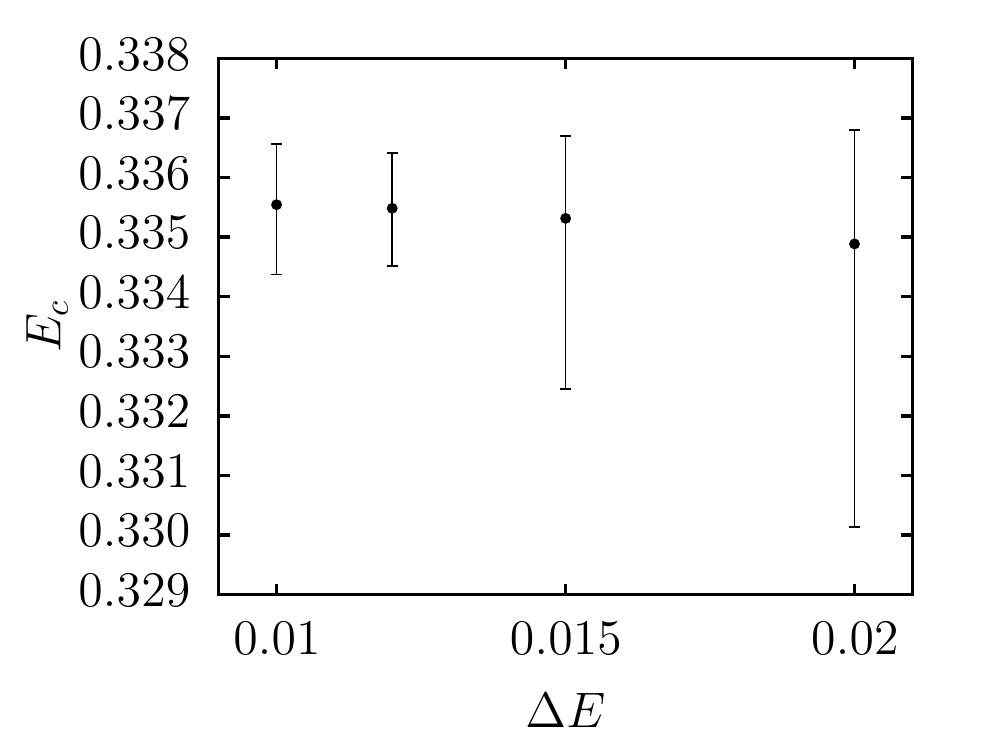}  \includegraphics[width=.49\columnwidth]{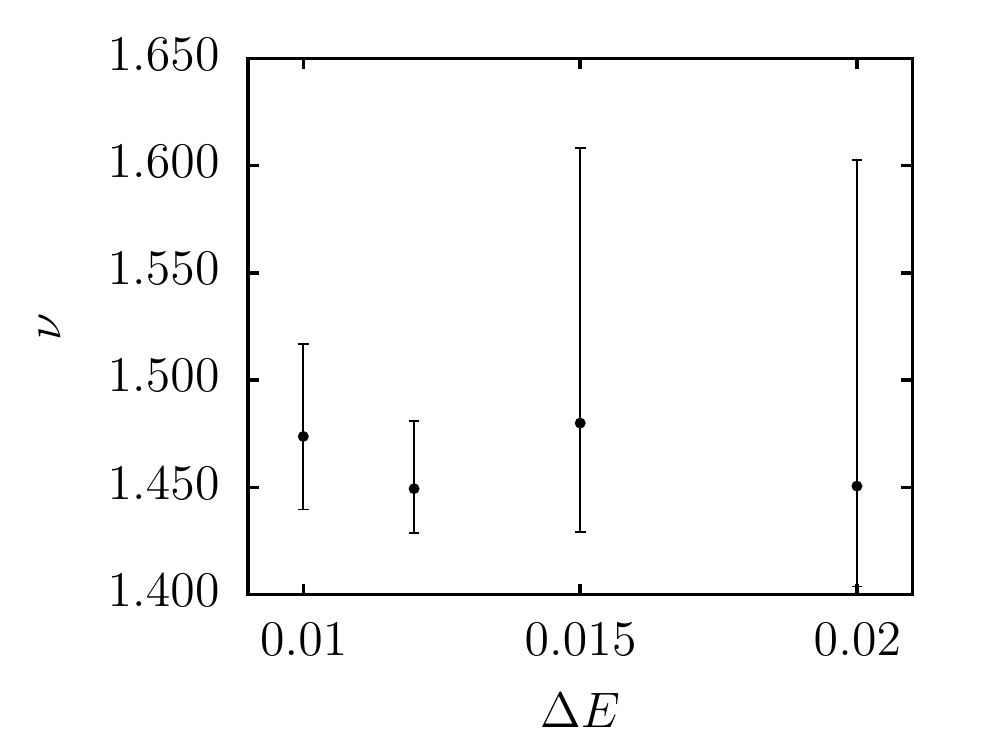}

  \caption{Dependence of the fitted critical point and critical
    exponent, as obtained from 
    $D_{0.1}^{ens}$ at fixed $\lambda=0.125$, for various energy
    windows $\Delta E$. Error bars correspond to the $95\%$ confidence
    band. Only statistical errors are shown.} 
  \label{fig:qcd_DDeltaE}
  \end{center}
\end{figure}

The resulting critical parameters are listed in
Tab.~\ref{tab:qcd_fitres_lambda0125} and shown in 
Fig.~\ref{fig:qcd_fitres_lambda0125}.
The results are essentially independent of $q$ and the type of
averaging, as expected. 
We also checked that the critical parameters do not depend on the
width of the energy window, $\Delta E$, used in the computation of the
GMFEs. As we show in Fig.~\ref{fig:qcd_DDeltaE}, the results for $E_c$
and $\nu$ are independent of $\Delta E$ within errors. Moreover, the
choice $\Delta E=0.0012$ is optimal, as it leads to the best accuracy.

\begin{table*}
	\begin {center}
	\begin{tabular}{|c| c| c |c| c| c | c| c| c| c| c|}
	\hline
	
{$\scriptstyle q$} & {\scriptsize exp} & $\scriptstyle{E_c}$ &
$\scriptstyle{\varepsilon^{\rm syst}_{E_c}}$ & $\scriptstyle{\nu}$ & $\scriptstyle{\varepsilon^{\rm syst}_{\nu}}$ &
$\scriptstyle{y}$ & $\scriptstyle{\varepsilon^{\rm syst}_{y}}$ &
$\scriptstyle{N_{df}}$ & $\scriptstyle{\chi^2}$ & $\scriptstyle{n_r
  n_{ir} n_\varrho n_\eta}$\\ \hline 
$\scriptstyle{0}$ &  $\scriptstyle{\alpha^{ens/typ}}$ &
$\scriptstyle{0.3353\ (0.3340..0.3363)}$ & $\scriptstyle{0.0004}$ &
$\scriptstyle{1.443\ (1.421..1.478)}$ & $\scriptstyle{0.056}$ &
$\scriptstyle{3.069\ (2.382..4.010)}$ & $\scriptstyle{0.278}$ &
$\scriptstyle{118}$ & $\scriptstyle{120}$ & $\scriptstyle{4\ 2\ 2\ 0}$ \\ 
 \hline
\multirow{2}{*}{$\scriptstyle{0.1}$} &  $\scriptstyle{D^{ens}}$ & 
$\scriptstyle{0.3355\ (0.3345..0.3364)}$ & $\scriptstyle{0.0003}$&
$\scriptstyle{1.449\ (1.429..1.481)}$ & $\scriptstyle{0.048}$&
$\scriptstyle{3.130\ (2.509..4.094)}$ & $\scriptstyle{0.240}$&
$\scriptstyle{118}$ & $\scriptstyle{119}$ & $\scriptstyle{4\ 2\ 2\ 0}$ \\ 
  &  $\scriptstyle{D^{typ}}$ & 
$\scriptstyle{0.3354\ (0.3344..0.3365)}$ & $\scriptstyle{0.0007}$&
$\scriptstyle{1.456\ (1.425..1.478)}$ & $\scriptstyle{0.048}$&
$\scriptstyle{3.322\ (2.564..4.301)}$ & $\scriptstyle{0.412}$&
$\scriptstyle{118}$ & $\scriptstyle{120}$ & $\scriptstyle{4\ 2\ 2\ 0}$ \\ 
 \hline
\multirow{4}{*}{$\scriptstyle{0.25}$} &  $\scriptstyle{\alpha^{ens}}$
& 
$\scriptstyle{0.3359\ (0.3342..0.3368)}$ & $\scriptstyle{0.0001}$ &
$\scriptstyle{1.470\ (1.437..1.521)}$ & $\scriptstyle{0.026}$ &
$\scriptstyle{3.380\ (2.217..4.683)}$ & $\scriptstyle{0.056}$ &
$\scriptstyle{118}$ & $\scriptstyle{118}$ & $\scriptstyle{4\ 2\ 2\ 0}$ \\ 
  &  $\scriptstyle{\alpha^{typ}}$ & 
$\scriptstyle{0.3358\ (0.3341..0.3365)}$ & $\scriptstyle{0.0001}$ &
$\scriptstyle{1.485\ (1.457..1.539)}$ & $\scriptstyle{0.026}$ &
$\scriptstyle{3.736\ (2.443..4.896)}$ & $\scriptstyle{0.148}$ &
$\scriptstyle{117}$ & $\scriptstyle{121}$ & $\scriptstyle{4\ 2\ 2\ 1}$ \\ 
  &  $\scriptstyle{D^{ens}}$ & 
$\scriptstyle{0.3355\ (0.3340..0.3366)}$ & $\scriptstyle{0.0002}$ &
$\scriptstyle{1.457\ (1.426..1.494)}$ & $\scriptstyle{0.048}$ &
$\scriptstyle{3.190\ (2.258..4.134)}$ & $\scriptstyle{0.188}$ &
$\scriptstyle{118}$ & $\scriptstyle{117}$ & $\scriptstyle{4\ 2\ 2\ 0}$ \\ 
  &  $\scriptstyle{D^{typ}}$ & 
$\scriptstyle{0.3354\ (0.3333..0.3362)}$ & $\scriptstyle{0.0004}$ &
$\scriptstyle{1.488\ (1.448..1.567)}$ & $\scriptstyle{0.054}$ &
$\scriptstyle{3.228\ (1.971..4.058)}$ & $\scriptstyle{0.334}$ &
$\scriptstyle{117}$ & $\scriptstyle{116}$ & $\scriptstyle{4\ 3\ 2\ 0}$ \\ 
 \hline
\multirow{2}{*}{$\scriptstyle{0.5}$} &  $\scriptstyle{D^{ens}}$ & 
$\scriptstyle{0.3357\ (0.3346..0.3369)}$ & $\scriptstyle{0.0001}$ &
$\scriptstyle{1.466\ (1.433..1.510)}$ & $\scriptstyle{0.040}$ &
$\scriptstyle{3.220\ (2.416..4.504)}$ & $\scriptstyle{0.118}$ &
$\scriptstyle{118}$ & $\scriptstyle{117}$ & $\scriptstyle{4\ 2\ 2\ 0}$ \\ 
  &  $\scriptstyle{D^{typ}}$ & 
$\scriptstyle{0.3356\ (0.3324..0.3368)}$ & $\scriptstyle{0.0001}$ &
$\scriptstyle{1.450\ (1.416..1.496)}$ & $\scriptstyle{0.036}$ &
$\scriptstyle{3.356\ (1.666..4.845)}$ & $\scriptstyle{0.148}$ &
$\scriptstyle{116}$ & $\scriptstyle{117}$ & $\scriptstyle{4\ 3\ 2\ 1}$ \\ 
 \hline
\multirow{4}{*}{$\scriptstyle{0.75}$} &  $\scriptstyle{\alpha^{ens}}$ & 
$\scriptstyle{0.3356\ (0.3339..0.3366)}$ & $\scriptstyle{0.0002}$ &
$\scriptstyle{1.462\ (1.424..1.517)}$ & $\scriptstyle{0.044}$ &
$\scriptstyle{3.221\ (2.154..4.364)}$ & $\scriptstyle{0.184}$ &
$\scriptstyle{118}$ & $\scriptstyle{119}$ & $\scriptstyle{4\ 2\ 2\ 0}$ \\ 
  &  $\scriptstyle{\alpha^{typ}}$ & 
$\scriptstyle{0.3355\  (0.3330..0.3366)}$ & $\scriptstyle{0.0001}$ &
$\scriptstyle{1.465\ (1.443..1.543)}$ & $\scriptstyle{0.032}$ &
$\scriptstyle{3.453\ (1.955..4.937)}$ & $\scriptstyle{0.194}$ &
$\scriptstyle{117}$ & $\scriptstyle{122}$ & $\scriptstyle{4\ 2\ 2\ 1}$ \\ 
  &  $\scriptstyle{D^{ens}}$ & 
$\scriptstyle{0.3361\ (0.3348..0.3371)}$ & $\scriptstyle{0.0001}$ &
$\scriptstyle{1.468\ (1.428..1.507)}$ & $\scriptstyle{0.038}$ &
$\scriptstyle{3.264\ (2.392..4.563)}$ & $\scriptstyle{0.118}$ &
$\scriptstyle{118}$ & $\scriptstyle{117}$ & $\scriptstyle{4\ 2\ 2\ 0}$ \\ 
  &  $\scriptstyle{D^{typ}}$ & 
$\scriptstyle{0.3360\ (0.3340..0.3371)}$ & $\scriptstyle{0.0001}$ &
$\scriptstyle{1.449\ (1.425..1.529)}$ & $\scriptstyle{0.034}$ &
$\scriptstyle{3.394\ (2.127..5.271)}$ & $\scriptstyle{0.130}$ &
$\scriptstyle{117}$ & $\scriptstyle{119}$ & $\scriptstyle{4\ 2\ 2\ 1}$ \\ 
 \hline
\multirow{2}{*}{$\scriptstyle{0.9}$} &  $\scriptstyle{D^{ens}}$ & 
$\scriptstyle{0.3363\ (0.3342..0.3374)}$ & $\scriptstyle{0.0002}$ &
$\scriptstyle{1.465\ (1.422..1.573)}$ & $\scriptstyle{0.036}$ &
$\scriptstyle{3.313\ (1.984..4.770)}$ & $\scriptstyle{0.128}$ &
$\scriptstyle{118}$ & $\scriptstyle{118}$ & $\scriptstyle{4\ 2\ 2\ 0}$ \\ 
  &  $\scriptstyle{D^{typ}}$ & 
$\scriptstyle{0.3361\ (0.3344..0.3372)}$ & $\scriptstyle{0.0002}$ &
$\scriptstyle{1.437\ (1.412..1.538)}$ & $\scriptstyle{0.036}$ &
$\scriptstyle{3.298\ (2.145..4.711)}$ & $\scriptstyle{0.256}$ &
$\scriptstyle{117}$ & $\scriptstyle{118}$ & $\scriptstyle{4\ 2\ 2\ 1}$ \\ 
 \hline
$\scriptstyle{1}$ &  $\scriptstyle{\alpha^{ens/typ}}$ & 
$\scriptstyle{0.3364\ (0.3346..0.3376)}$ & $\scriptstyle{0.0001}$ &
$\scriptstyle{1.464\ (1.425..1.535)}$ & $\scriptstyle{0.034}$ &
$\scriptstyle{3.334\ (2.175..5.018)}$ & $\scriptstyle{0.108}$ &
$\scriptstyle{118}$ & $\scriptstyle{118}$ & $\scriptstyle{4\ 2\ 2\ 0}$ \\ 
 \hline
  \end{tabular}
  \caption{Result of the MFSS at fixed $\lambda=0.125$ for the eigenvectors of the Dirac operator of QCD.}
  \label{tab:qcd_fitres_lambda0125}

  \end{center}
\end{table*}

%
%
%
%



To quote a final result for the critical parameters, we have averaged
the values of $E_c$, $\nu$ and $y$, and of the corresponding errors, 
obtained with the various GMFEs. (A weighted average, using the
inverse of the error band as weight, yields similar numbers.)
Our result for the critical point, $E_c=0.3357\ (0.3340..0.3368)$,
is compatible with the value reported in Ref.~\onlinecite{Giordano14}
at the 2-$\sigma$ level. On average, the systematic error on $E_c$ is
$\varepsilon_{E_c}^{\rm syst}=0.0002$, so negligible compared to the
statistical error. Our result for the critical exponent, $\nu=1.461\
(1.429..1.519)$, agrees at the 1-$\sigma$ level with the result of
Ref.~\onlinecite{Giordano14}, and with previous results for the
critical exponent of the unitary Anderson model~\cite{Ujfalusi15,SO}. 
For this quantity, one has also to take into account
that on average the systematic error due to truncation,
$\varepsilon_{\nu}^{\rm syst}=0.040$, is of the same size as the 
statistical error. 
On the other hand, our value for the irrelevant exponent, $y=3.307\
(2.210..4.572)$, is significantly different from 
the value of Ref.~\onlinecite{Ujfalusi15}, $y^{\rm UV}=1.651\
(1.601..1.707)$. It is well known that it is very difficult to
determine irrelevant exponents accurately, and to explain this
discrepancy further work and higher-quality data are needed. 
It is possible that for the system sizes presently available, 
more than one irrelevant term gives important contributions, so that
our result for $y$ would be a sort of ``effective'' irrelevant
exponent. In any case this point requires further analysis.

As a final remark, we note that since results for different $q$ are
strongly correlated, there is no significance in the fact that our
values for the critical point are systematically lower, and the ones
for the critical exponent are systematically higher than the reference
values. 

\begin{figure*}[t!]
  \begin {center}
  \begin{tabular}{c c}
  \begin{overpic}[type=pdf,ext=.pdf,read=.pdf,width=.45\textwidth]{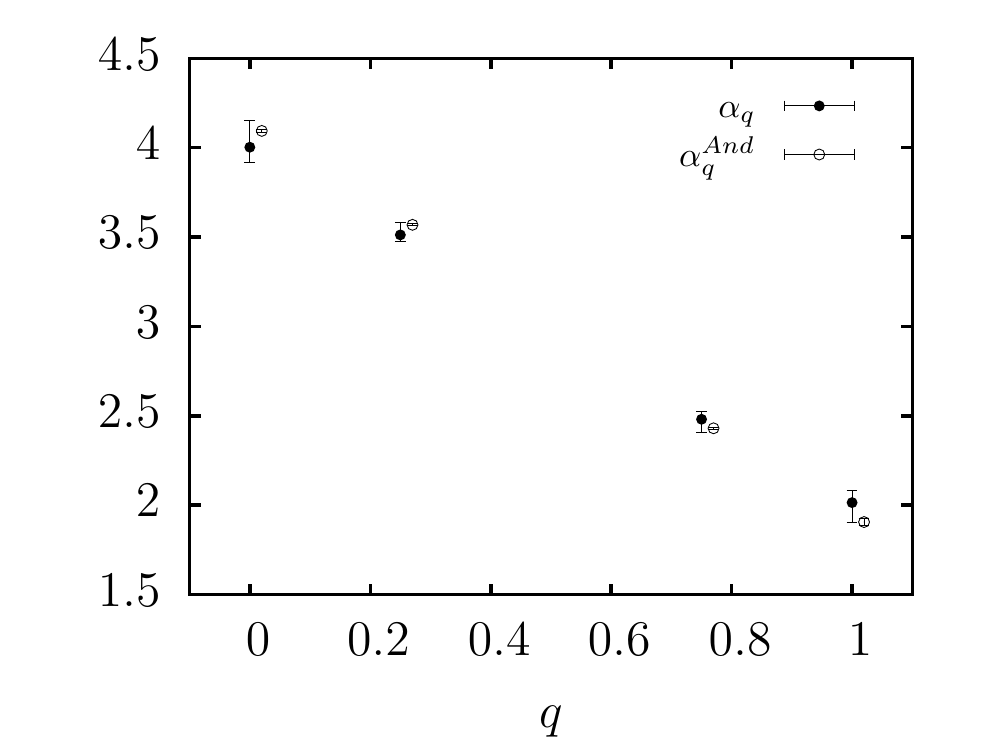}
		\put(0,74){\large (a)}
	\end{overpic} &
  \begin{overpic}[type=pdf,ext=.pdf,read=.pdf,width=.45\textwidth]{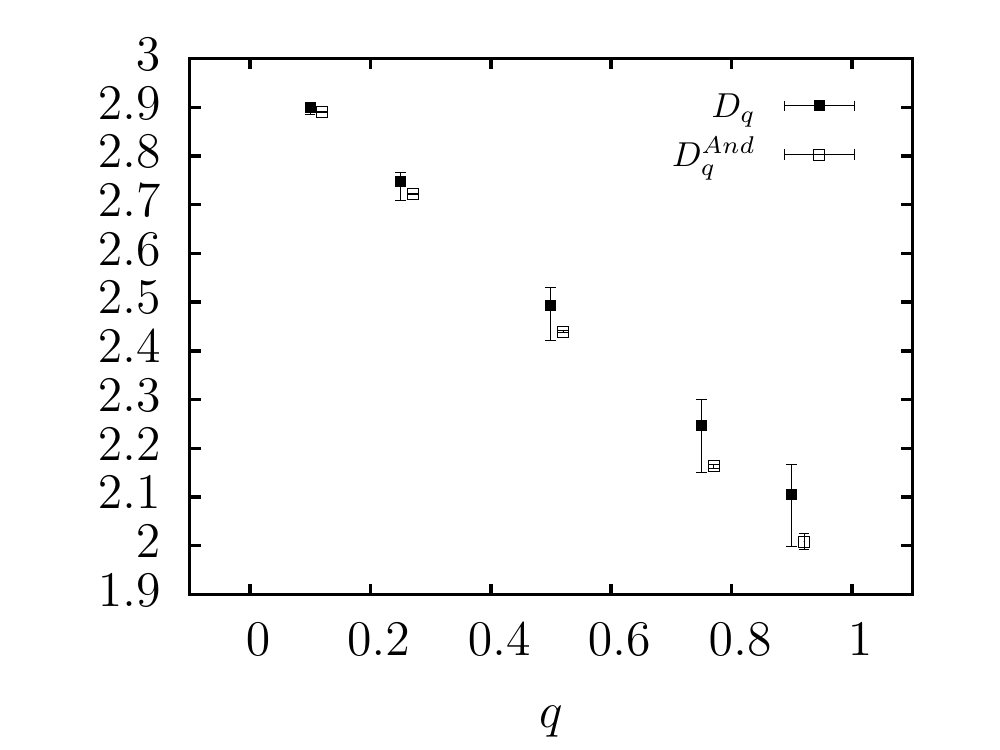} 
		\put(0,74){\large (b)}
	\end{overpic}\\
  \end{tabular}
  \caption{Estimated values of the MFEs, (a) $\alpha_q$ and (b) $D_q$,
    in high-temperature QCD, and MFEs of the 3D unitary Anderson model
    taken from Ref.~\onlinecite{Ujfalusi15} (slightly shifted
    horizontally for clarity). 
} 
  \label{fig:qcd_fitres_MFEs}
  \end{center}
\end{figure*}

The convergence of the fixed-$\lambda$ MFSS confirms the presence of
a critical point in the QCD Dirac spectrum where the system undergoes
a true localization-delocalization transition, employing completely
different observables than the ones used in
Ref.~\onlinecite{Giordano14}. The results of our analysis also provide 
further evidence that the transition in the QCD Dirac spectrum belongs
to the universality class of the 3D unitary Anderson model.
Moreover, despite the fact that it does not provide the values of the
MFEs, the convergence of this method also strongly indicates the
presence of multifractality at the critical point. 

We next procedeed to apply the variable-$\lambda$ method, in order
to try and determine the multifractal exponents, and compare them to
the ones obtained for the unitary Anderson model. However, this method
requires small values of $\lambda$ to work properly, and is also more
demanding as it is a two-variable fit. In practice, the
$\chi^2/N_{df}$ ratio reached a value close to unity only if we left
out the smallest system sizes, below $L_{min}=36$, and if we used data
corresponding to $\ell=1$ and $2$ only. Although using $L_{min}=36$
and $\ell=1,2$ improved the convergence, the fits were still unstable
against changing the expansion orders. This can be understood, as a
similar amount of independent data is available as in the
fixed-$\lambda$ method, but there are many more parameters to fit, as
discussed in Sec.~\ref{sec:MFSS}. In order to be able to estimate the
MFEs, we then fixed the critical energy and the critical exponent to
the values obtained with the fixed-$\lambda$ method, $E_c=0.3357$ and
$\nu=1.461$, in this way stabilizing the fits. The systematic
uncertainty corresponding to this procedure was estimated by repeating
the fits with $E_c$ and $\nu$ fixed to one of 
the four possible combinations of the values $E_c^{l,u}$ and
$\nu^{l,u}$, which are the lower and upper boundaries of the
confidence interval of $E_c$ and $\nu$, respectively. 
The largest and smallest values obtained in this way were then used as
upper and lower error bar on the MFEs. We experienced that the main
source of uncertainty comes from the choice of $E_c$, while fits are
much less sensitive to the choice of $\nu$. Moreover, statistical
errors (estimated by Monte-Carlo) and systematic errors due to
truncation were comparatively negligible. 

The results of this procedure are depicted in
Fig.~\ref{fig:qcd_fitres_MFEs}. A set of nontrivial 
MFEs was obtained, thus providing direct evidence of the
multifractality of the critical eigenfunctions of the QCD Dirac
operator. Moreover, our results for the MFEs in QCD are compatible
with the ones obtained in the unitary Anderson model, which further
confirms that the transition belongs to the chiral unitary Anderson
class.

\section{Summary}
\label{sec:concl}
We investigated the Anderson transition in the spectrum of the Dirac
operator of QCD at high temperature, found by the authors of
Ref.~\onlinecite{Giordano14}. While that work made use of spectral
statistics, our aim in this paper was to examine the transition
by studying the eigenvectors, and their multifractal properties at the
critical energy. The results of Ref.~\onlinecite{Giordano14} for the
correlation length critical exponent suggested that the Anderson
transition in QCD belongs to the same universality class as the
three-dimensional unitary Anderson model. We therefore looked for more
similarities between these models.

First we examined the correlations between
eigenvectors of a given gauge configuration. We found strong
correlations between eigenmodes of the QCD Dirac operator, decreasing with
energy separation in a similar way as in the unitary Anderson model.
We then performed two multifractal finite-size scaling (MFSS) analyses, one
with fixed ratio $\lambda$ of the coarse-graining box size to the
system size, and one with variable $\lambda$. 
MFSS with the fixed-$\lambda$ method allowed an alternative
determination of the critical point and of the critical exponent,
which is in agreement with the findings of
Ref.~\onlinecite{Giordano14}, and, for the critical exponent, with
those of Refs.~\onlinecite{Ujfalusi15,SO} for the unitary Anderson
model. To perform MFSS with the variable-$\lambda$ method and
determine the multifractal exponents (MFEs), we performed fits fixing
the critical energy and the critical exponent to the values obtained
with the fixed-$\lambda$ method. The resulting MFEs are compatible
with the MFEs found in the unitary Anderson model. 

In conclusion, our work confirms the presence of an Anderson
metal-insulator phase transition in the spectrum of the Dirac operator
in high-temperature QCD, and provides further evidence that this
transition belongs to the three-dimensional unitary Anderson model
class. Morever, we have shown that the critical wave-functions of the
Dirac operator are multifractals. 
The physical consequences of the QCD
Anderson transition and of multifractality still largely need to be
explored, and may lead in particular to a better understanding of the
QCD chiral transition. Further work along these lines might prove
beneficial for condensed matter physics as well, as it approaches the
subject of localization/delocalization transitions from a broader
perspective.


\begin{acknowledgments}
  We thank the Budapest-Wuppertal group for allowing us to use their code to
  generate the gauge configurations.
  Financial support to LU and IV from OTKA under Grant No. K108676,
  and from the Alexander von Humboldt Foundation is gratefully acknowledged. 
  MG and TGK are supported by the Hungarian Academy of Sciences under
  ``Lend\"ulet'' grant No. LP2011-011. FP is supported by OTKA under the 
  grant OTKA-NF-104034. 

\end{acknowledgments}

\end{document}